\documentclass[a4paper]{VulcanTechReport}

\usepackage{cite}
\usepackage{tikz}
\usetikzlibrary{shapes.geometric,arrows.meta,positioning,calc,backgrounds}
\usepackage{pgfplots}
\pgfplotsset{compat=1.18}
\usepackage{amsmath}
\usepackage{amssymb}
\usepackage{adjustbox}
\usepackage{multirow}
\usepackage{subcaption}
\usepackage{colortbl}

\definecolor{vulcanyellow}{RGB}{255,202,58}
\definecolor{ttcolor}{HTML}{1A1A1A}
\definecolor{codebg}{HTML}{F0F0F0}
\hypersetup{urlcolor=ttcolor}

\usepackage[most]{tcolorbox}
\newtcbox{\code}{on line,
  boxrule=0pt, boxsep=0pt,
  top=2pt, bottom=2pt, left=3pt, right=3pt,
  arc=3pt,
  colback=codebg, coltext=ttcolor,
  fontupper=\ttfamily}
\newcommand{\icode}[1]{{\color{ttcolor}\ttfamily #1}}

\reporttitle{MCPThreatHive: Automated Threat Intelligence for Model Context Protocol Ecosystems}
\reportsubtitle{Vulcan Research, AIFT}
\reportauthors{Yi Ting Shen, Kentaroh Toyoda, and Alex Leung}
\reportdate{April 2026}

\leftheadercontent{MCPThreatHive}
\rightheadercontent{\includegraphics[width=3cm]{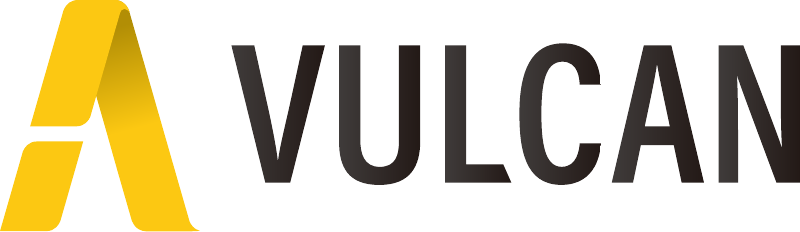}}

\begin{document}


\thispagestyle{empty}
\vspace*{-0.075\textheight}
\hfill\includegraphics[width=5cm]{vulcan-logo.pdf}
\vspace{0.06\textheight}

{\large\raggedright\reportdate\par}
\vspace{0.01\textheight}

{\fontsize{32pt}{34pt}\selectfont\raggedright\textbf{\reporttitle}\par}
\vspace{0.03\textheight}

{\Large\raggedright\textit{\textbf{\reportsubtitle}}\par}
\vspace{0.04\textheight}

{\large\raggedright\reportauthors\par}
\vspace{0.04\textheight}

{\normalsize\raggedright\url{https://github.com/VulcanLab/MCPThreatHive}\par}
\vspace{0.06\textheight}

\tableofcontents


\begin{abstract}
\noindent
The rapid proliferation of Model Context Protocol (MCP)-based agentic systems has introduced a new category of security threats that existing frameworks are inadequately equipped to address. We present MCPThreatHive, an open-source platform that automates the end-to-end lifecycle of MCP threat intelligence: from continuous, multi-source data collection through AI-driven threat extraction and classification, to structured knowledge graph storage and interactive visualization. The platform operationalizes the MCP-38 threat taxonomy, a curated set of 38 MCP-specific threat patterns mapped to STRIDE, OWASP Top~10 for LLM Applications, and OWASP Top~10 for Agentic Applications. A composite risk scoring model provides quantitative prioritization. Through a comparative analysis of representative existing MCP security tools, we identify three critical coverage gaps that MCPThreatHive addresses: incomplete compositional attack modeling, absence of continuous threat intelligence, and lack of unified multi-framework classification.
\end{abstract}

\medskip\noindent\rule{\linewidth}{0.4pt}\medskip


\section{Introduction}
\label{sec:introduction}

The Model Context Protocol (MCP), introduced by Anthropic in late 2024~\cite{mcp-spec-2024}, has rapidly emerged as the de facto standard for connecting large language model (LLM)-based agents to external tools and data sources. By providing a standardized, bidirectional communication channel between AI host applications and tool servers, MCP has accelerated the development of autonomous agentic workflows spanning file systems, databases, web services, and code execution environments. As of early 2026, hundreds of open-source and commercial MCP servers are publicly available, and major AI platforms, including Claude, GitHub Copilot, and Cursor, have adopted MCP as a first-class integration mechanism.

This rapid adoption has outpaced the security community's ability to characterize and mitigate the associated risks. Unlike traditional software interfaces, MCP tool selection and invocation are mediated by natural-language descriptions interpreted by an LLM. This semantic mediation introduces attack vectors (tool description poisoning, indirect prompt injection via external data, parasitic tool chaining) with no direct analogues in classical threat models. Concurrently, agentic systems operate with significant autonomy, accumulate rich contextual state across multi-turn interactions, and may invoke dozens of tools per session, dramatically increasing the blast radius of a successful compromise.

Existing security frameworks and tools address fragments of this problem but leave critical gaps. A comparative analysis of representative MCP security tools (Section~\ref{sec:eval-comparison}) reveals three recurring shortcomings: incomplete compositional attack modeling, absence of continuous threat intelligence, and lack of unified multi-framework classification.

We present MCPThreatHive, an open-source platform that automates the end-to-end lifecycle of MCP threat intelligence: from continuous, multi-source data collection through AI-driven threat extraction and classification, to knowledge graph construction and interactive visualization. The platform operationalizes the MCP-38 threat taxonomy~\cite{shen2026mcp38}, a curated set of 38 MCP-specific threat patterns mapped to STRIDE, OWASP Top~10 for LLM Applications~\cite{owasp-llm-2023, owasp-llm-2025}, and OWASP Top~10 for Agentic Applications~\cite{owasp-agentic-2024}. A composite risk scoring model adapted from DREAD~\cite{shostack2014threat} provides quantitative prioritization.

The key contributions of this paper are:
\begin{enumerate}
    \item An automated four-stage pipeline for continuous MCP threat intelligence, covering collection, AI-powered analysis, knowledge graph storage, and interactive visualization.
    \item A composite risk scoring model combining security impact severity, attack success rate, persistence scope, and exploitation ease, with MCP-specific priority multipliers.
    \item A comparative analysis of representative existing MCP security tools, identifying three critical coverage gaps that MCPThreatHive addresses.
    \item An open-source implementation released for community use and extension.
\end{enumerate}

The remainder of this paper is organized as follows. Section~\ref{sec:background} surveys the MCP ecosystem and related work. Section~\ref{sec:problem} presents the problem statement and gap analysis. Section~\ref{sec:mcpthreathive} describes the MCPThreatHive platform. Section~\ref{sec:evaluation} compares MCPThreatHive with existing tools and evaluates the pipeline against a known incident. Section~\ref{sec:discussion} discusses limitations, and Section~\ref{sec:conclusion} concludes.


\section{Background and Related Work}
\label{sec:background}

This section surveys the MCP ecosystem, known attack classes, and the security frameworks and tools that have been proposed to address them. We identify the specific limitations of each, which together motivate the design of MCPThreatHive.

\subsection{The Model Context Protocol}
\label{sec:bg-mcp}

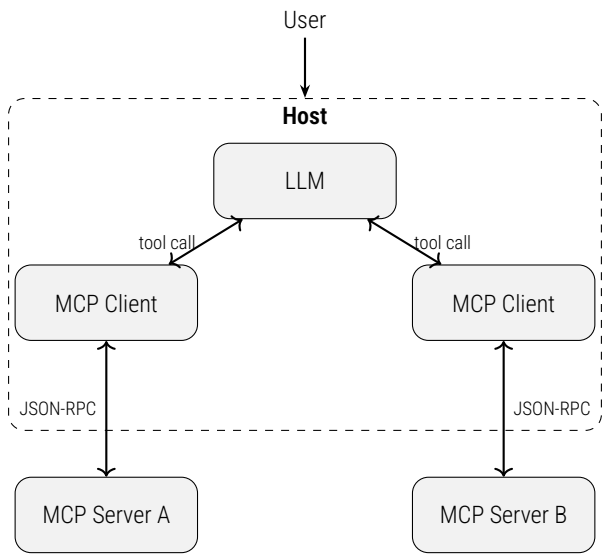
\begin{figure}[tbp]
\centering
\begin{tikzpicture}[
    box/.style={draw, rounded corners=6pt, minimum width=2.4cm, minimum height=1cm, align=center, font=\small},
    arr/.style={-{Stealth[length=5pt]}, thick},
    lbl/.style={font=\scriptsize, midway, align=center},
    node distance=1.8cm
]

\node[font=\small] (user) {User};

\node[box, minimum width=7.8cm, minimum height=4.4cm, dashed, below=0.8cm of user] (host) {};
\node[font=\small\bfseries, anchor=north] at (host.north) {Host};

\node[box, fill=gray!10] (llm) at ([yshift=-1.1cm]host.north) {LLM};

\node[box, fill=gray!10, below left=0.6cm and 0.2cm of llm] (client1) {MCP Client};
\node[box, fill=gray!10, below right=0.6cm and 0.2cm of llm] (client2) {MCP Client};

\node[box, fill=gray!10, below=1.8cm of client1] (server1) {MCP Server A};
\node[box, fill=gray!10, below=1.8cm of client2] (server2) {MCP Server B};

\draw[arr] (user) -- (host);
\draw[arr, <->] (llm) -- node[lbl, left] {tool call} (client1);
\draw[arr, <->] (llm) -- node[lbl, right] {tool call} (client2);
\draw[arr, <->] (client1) -- node[lbl, left] {JSON-RPC} (server1);
\draw[arr, <->] (client2) -- node[lbl, right] {JSON-RPC} (server2);

\end{tikzpicture}
\caption{MCP three-tier architecture. The Host embeds an LLM and one or more MCP Clients, each maintaining a stateful session with an MCP Server over JSON-RPC~2.0.}
\label{fig:mcp-arch}
\end{figure}

The Model Context Protocol (MCP) is an open standard that defines how AI host applications communicate with external tool servers~\cite{mcp-spec-2024}. The architecture comprises three roles: the \emph{Host} (user-facing application with an embedded LLM), one or more \emph{MCP Clients} (protocol handlers that maintain stateful sessions), and \emph{MCP Servers} (lightweight processes that expose tools, resources, and prompts via JSON-RPC~2.0). Communication occurs over either local \icode{stdio} pipes or remote HTTP with Server-Sent Events (SSE). Figure~\ref{fig:mcp-arch} illustrates this three-tier architecture.

The defining characteristic of MCP is that tool selection is delegated to the LLM at inference time. The LLM reads natural-language \icode{description} fields from server-provided tool manifests and decides which tool to invoke and with which parameters. This design enables flexible, context-sensitive tool use but simultaneously introduces a \emph{semantic attack surface}: any actor who can write arbitrary text into a field the LLM reads (tool descriptions, resource content, external documents) can influence the model's decisions without altering any execution logic.

Each tool manifest contains \icode{name}, \icode{description}, \icode{inputSchema} (a JSON Schema), and optional \icode{annotations}. Since the \icode{description} field is unconstrained free text interpreted by the LLM, it constitutes the primary injection surface. MCP currently lacks mandatory cryptographic signing, content-addressing, or version pinning for tool manifests, meaning that a server audited at time $t_0$ may exhibit different behavior at time $t_1$.

\subsection{Known MCP Attack Classes}
\label{sec:bg-attacks}

Published research has identified several concrete attack classes since MCP's introduction in late 2024. Shen et al.~\cite{shen2026mcp38} systematically catalog these into 38 protocol-specific threat patterns (MCP-01 through MCP-38) organized across five risk categories; we summarize the most prominent classes below.

\paragraph{Indirect Prompt Injection (IPI)}
Attackers embed malicious instructions in external data consumed by MCP tools, such as web pages, documents, or emails, causing the agent to execute unintended commands. Greshake et al.~\cite{greshake2023prompt} first demonstrated IPI in LLM-integrated applications; subsequent work shows that MCP-connected agents are significantly more susceptible due to their autonomous tool-invocation capability and wider data exposure surface~\cite{yang2025mcpsecbench, zhang2025mcpsecuritybenchmsb}.

\paragraph{Parasitic Tool Chains (MCP-UPD)}
Zhao et al.~\cite{zhao2025mindserver} formalise the MCP Unintended Privacy Disclosure (UPD) model, in which individually benign tools are chained to exfiltrate data or escalate privileges. For example, a \icode{weather\_api} call returns output embedded with a malicious payload, which is then written to disk by a \icode{file\_write} tool. Each individual tool is legitimate; the attack emerges from their composition under adversarial direction.

\paragraph{Preference Manipulation (MPMA)}
Wang et al.~\cite{wang2025mpma} demonstrate that crafting specific metadata fields, such as priority annotations or LLM-preference signals in tool descriptions, can statistically bias the LLM's tool selection toward attacker-controlled tools, even when trusted alternatives are available.

\paragraph{Tool Description Poisoning}
Malicious tool descriptions can embed hidden instructions using Unicode zero-width characters, homoglyphs, or strategically placed natural-language commands that are invisible to human reviewers but influence LLM decision-making. Full schema poisoning extends this technique to the JSON Schema \icode{description} fields within \icode{inputSchema}, affecting how the model generates invocation parameters~\cite{yang2025mcpsecbench}.

\paragraph{Rug Pull / Dynamic Mutation}
MCP server operators can update server behavior after users have established trust. Because MCP lacks cryptographic content-addressing or version pinning for tool descriptions, a server that was audited at time $t_0$ may exhibit entirely different behavior at time $t_1$, analogous to a supply chain attack operating at the semantic layer.

\paragraph{Real-world incidents}
These attack classes are not purely theoretical. The GitHub MCP vulnerability (2025) demonstrated indirect prompt injection via crafted repository content, enabling exfiltration of private data from other users' sessions~\cite{github-mcp-vuln-2025}. CVE-2025-6514~\cite{cve-2025-6514} exposed an OS command injection vulnerability in mcp-remote, a widely used MCP transport library, enabling remote code execution when connecting to a malicious server.

\subsection{Existing Security Frameworks}
\label{sec:bg-frameworks}

Several security frameworks provide partial coverage of MCP threats, but none was designed for the protocol's specific characteristics. Classical frameworks such as STRIDE~\cite{shostack2014threat} and MITRE ATT\&CK (including its AI supplement, ATLAS~\cite{mitre-atlas}) model threats in terms of deterministic system behavior and concrete system actions (e.g., file access, network connections). They do not natively capture semantic attacks where the adversary influences model \emph{reasoning} rather than system \emph{logic}. Nevertheless, STRIDE's six categories remain useful as a classification backbone; the MCP-38 taxonomy~\cite{shen2026mcp38} maps all 38 MCP threats to their primary STRIDE categories, and MCPThreatHive operationalizes this mapping for automated classification.

The OWASP Top~10 for LLM Applications~\cite{owasp-llm-2023, owasp-llm-2025} and the OWASP Top~10 for Agentic Applications~\cite{owasp-agentic-2024} advance coverage toward AI-specific risks such as prompt injection (LLM01) and agent goal hijacking (ASI01). However, both operate at a high abstraction level and do not address the compositionality of multi-tool MCP workflows or protocol-level threats such as transport tampering and manifest poisoning.

The closest existing work to a protocol-specific framework is MCPSecBench~\cite{yang2025mcpsecbench}, which provides a 4$\times$17 benchmark matrix covering four attack surfaces against 17 attack categories. The MCP-38 taxonomy~\cite{shen2026mcp38} cross-maps all 38 threat patterns to this matrix. However, MCPSecBench is a benchmarking tool rather than a continuous threat intelligence platform and does not provide automated classification, knowledge graph construction, or risk scoring. Similarly, the Cloud Security Alliance's MAESTRO framework~\cite{maestro2025} offers useful architectural decomposition across seven layers for agentic AI systems but operates at the system design level and lacks automated threat detection.

\subsection{Threat Intelligence Platforms and MCP Security Tools}
\label{sec:bg-tools}

Established threat intelligence platforms such as MISP~\cite{wagner2016misp} and OpenCTI~\cite{opencti2023} provide collaborative sharing and structured storage for indicators of compromise (IoCs), but they are designed for network and endpoint threats. They lack native support for the semantic, inference-time threat classes that characterize MCP ecosystems.

Within the MCP-specific tool ecosystem, several point solutions have emerged. MCP-Scan~\cite{mcp-scan-2025} performs static analysis of tool descriptions to detect known poisoning patterns. Other tools focus on runtime guardrails (e.g., Ramparts, MCP-Guardian), permission enforcement (e.g., Pangea MCP Proxy), or configuration auditing (e.g., Agentic Radar). While valuable for their respective use cases, these tools share three limitations: (1)~they operate on individual tool descriptions or invocations without modeling compositional attacks across tool chains, (2)~they provide point-in-time scanning rather than continuous threat intelligence, and (3)~they classify threats against a single framework rather than providing unified multi-framework coverage.

While the MCP-38 taxonomy~\cite{shen2026mcp38} provides the definitional foundation for MCP-specific threats, no existing platform combines continuous multi-source intelligence gathering, automated AI-driven threat classification, structured knowledge graph storage, and multi-framework mapping (STRIDE, OWASP LLM, OWASP Agentic, MCPSecBench, MAESTRO) into an integrated threat intelligence pipeline for MCP ecosystems. MCPThreatHive fills this gap by operationalising MCP-38 within an end-to-end automated platform.


\section{Problem Statement}
\label{sec:problem}

The previous section established that MCP introduces a structurally distinct attack surface and that existing frameworks and tools provide only partial coverage. This section defines the specific gaps that MCPThreatHive addresses; a detailed capability comparison with existing tools is presented in Section~\ref{sec:eval-comparison}.

Three critical gaps motivate this work. First, existing tools model threats in isolation and cannot detect compositional attacks where the risk emerges from the interaction between individually benign tools. Second, all reviewed tools operate in manual or point-in-time mode rather than continuously monitoring intelligence sources for newly emerging threats. Third, per-tool scanners classify threats against a single framework (e.g., STRIDE or OWASP LLM) and do not provide unified multi-framework coverage spanning STRIDE, OWASP LLM, OWASP Agentic, and MCP-38.


MCPThreatHive addresses these gaps through an integrated platform that provides: (1)~operationalisation of the MCP-38 threat taxonomy~\cite{shen2026mcp38} with composite risk scoring and multi-framework cross-mapping, (2)~continuous, automated intelligence gathering from diverse sources (CVE databases, security blogs, academic preprints, GitHub Security Advisories) with AI-driven threat extraction and classification, and (3)~a neuro-symbolic knowledge graph that captures compositional relationships between threats, tools, attack chains, and mitigations.


\section{MCPThreatHive}
\label{sec:mcpthreathive}

The gaps identified in the previous section call for a platform that goes beyond point-in-time scanning or single-framework classification. MCPThreatHive is our response: an open-source, end-to-end threat intelligence platform that continuously monitors the MCP security landscape, automatically classifies emerging threats against the MCP-38 taxonomy~\cite{shen2026mcp38} and multiple established frameworks, constructs a knowledge graph of threat relationships, and generates actionable risk plans. This section describes the platform's architecture, its core processing stages, and the key design decisions behind each component.

\subsection{System Architecture}
\label{sec:architecture}

\begin{figure*}[tbp]
\centering
\resizebox{\textwidth}{!}{%
\begin{tikzpicture}[
    stage/.style={draw, rounded corners=6pt, minimum width=3.0cm, minimum height=1.8cm, align=center, font=\small, fill=gray!10},
    source/.style={draw, rounded corners=4pt, minimum width=1.8cm, minimum height=0.7cm, align=center, font=\scriptsize, fill=white},
    arr/.style={-{Stealth[length=5pt]}, thick},
    lbl/.style={font=\scriptsize, midway, above},
    node distance=1.6cm
]

\node[source] (src1) {RSS Feeds};
\node[source, below=0.3cm of src1] (src2) {Web Search};
\node[source, below=0.3cm of src2] (src3) {NVD API};
\node[source, below=0.3cm of src3] (src4) {GitHub API};

\node[stage, right=1.4cm of $(src2.east)!0.5!(src3.east)$] (s1) {Intelligence\\Gathering};

\draw[arr] (src1.east) -- (s1);
\draw[arr] (src2.east) -- (s1);
\draw[arr] (src3.east) -- (s1);
\draw[arr] (src4.east) -- (s1);

\node[stage, right=1.8cm of s1] (s2) {AI Threat\\Analysis};

\node[font=\scriptsize, below=0.15cm of s2, align=center, text width=3.0cm] {CoT classification\\Risk scoring\\Output repair};

\node[stage, right=1.8cm of s2] (s3) {Structured\\Storage};

\node[font=\scriptsize, below=0.15cm of s3, align=center, text width=3.0cm] {SQL database\\Knowledge graph};

\node[stage, right=1.8cm of s3] (s4) {Visualization \&\\Risk Planning};

\node[font=\scriptsize, below=0.15cm of s4, align=center, text width=3.0cm] {Threat matrix\\3D landscape\\Risk planner};

\draw[arr] (s1) -- node[lbl] {IntelItems} (s2);
\draw[arr] (s2) -- node[lbl] {Threat cards} (s3);
\draw[arr] (s3) -- node[lbl] {Queries} (s4);

\node[font=\scriptsize\bfseries, anchor=south west] at (s1.north west) {1};
\node[font=\scriptsize\bfseries, anchor=south west] at (s2.north west) {2};
\node[font=\scriptsize\bfseries, anchor=south west] at (s3.north west) {3};
\node[font=\scriptsize\bfseries, anchor=south west] at (s4.north west) {4};

\end{tikzpicture}%
}
\caption{MCPThreatHive four-stage pipeline architecture. Data flows left to right: intelligence sources feed into collection, items are classified by an LLM, results are stored in a relational database and knowledge graph, and exposed through interactive visualization and risk planning interfaces.}
\label{fig:pipeline}
\end{figure*}
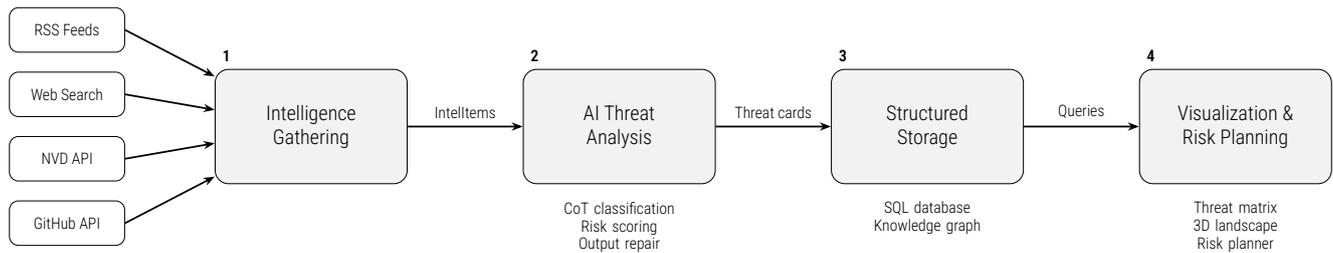

MCPThreatHive is structured as a four-stage pipeline (Figure~\ref{fig:pipeline}): (1)~intelligence gathering, (2)~AI-powered threat analysis, (3)~structured storage in a relational database and knowledge graph, and (4)~interactive visualization and risk planning.

\subsubsection{Intelligence Gathering}
\label{sec:arch-intel}

The platform collects threat intelligence from four complementary source types: web search via DuckDuckGo\footnote{\url{https://duckduckgo.com}}, a privacy-preserving search engine; RSS feeds (ArXiv CS.CR\footnote{\url{http://export.arxiv.org/rss/cs.CR}}, ArXiv CS.AI\footnote{\url{http://export.arxiv.org/rss/cs.AI}}, Krebs on Security\footnote{\url{https://krebsonsecurity.com/feed/}}, The Hacker News\footnote{\url{https://feeds.feedburner.com/TheHackersNews}}, and Schneier on Security\footnote{\url{https://www.schneier.com/feed/atom/}}); the NIST National Vulnerability Database (NVD) REST API\footnote{\url{https://services.nvd.nist.gov/rest/json/cves/2.0}}; and GitHub Security Advisories via the REST API\footnote{\url{https://api.github.com/advisories}}. All sources are normalized to a uniform \emph{IntelItem} schema comprising: identifier, title, content, source URL, source type, collection timestamp, and AI-assigned relevance score. This normalization allows all downstream processing to be source-agnostic.

Query generation is itself LLM-driven. An AI Keyword Generator produces targeted search queries seeded with the MCP-38 threat taxonomy, at multiple specificity levels, from broad (e.g., ``MCP security vulnerabilities'') to narrow (e.g., ``MCP tool description poisoning in Claude Desktop''). An iterative refinement loop automatically relaxes queries that return fewer than a minimum result threshold, preventing coverage gaps for low-frequency threat topics.

\subsubsection{AI Threat Analysis}
\label{sec:arch-analysis}

Intelligence items that pass a configurable relevance threshold ($\theta = 0.70$ by default) enter the threat analysis stage, which processes items in batches of three to five through a large language model via a provider-agnostic API~\cite{litellm2023}. To prevent context truncation during the extensive chain-of-thought reasoning required by frontier models, the platform allocates a fixed, generous output budget of up to 12{,}000 tokens for batch threat classification. This budget addresses a critical operational property of frontier reasoning models: these models perform internal chain-of-thought deliberation before producing output, consuming a substantial fraction of the allocated budget in intermediate reasoning steps. Allocating fewer tokens causes truncation of the output payload.

The system prompt enforces a chain-of-thought (CoT) analysis sequence before the model emits structured output: (1)~classify the MCP workflow phase (Task Planning, Tool Invocation, Response Handling, or Cross-Phase), (2)~assign threat identifiers from the MCP-01 to MCP-38 taxonomy, (3)~determine the primary STRIDE category, (4)~assess DREAD-adapted risk factors, and (5)~verify consistency between the textual risk level and the numeric risk score. Schema-aware prompting embeds semantic consistency constraints alongside the output JSON schema. For example, the \emph{Critical} risk level may be assigned only when the numeric score exceeds 9.0 and the threat involves remote code execution, data exfiltration, or critical asset compromise. This constraint injection reduces risk-level hallucination compared to unconstrained generation. Following this CoT classification, the system applies a deterministic taxonomy bridge: a lookup table maps each assigned MCP-38 identifier to its corresponding categories in the OWASP LLM Top~10 and OWASP Agentic Top~10 frameworks. This hybrid design intentionally separates semantic reasoning (LLM-assigned STRIDE and MCP-38 labels) from structured framework alignment (rule-based OWASP mapping), eliminating the risk of hallucinated framework assignments.

Because LLM-generated structured output may be syntactically malformed due to context-length truncation or grammar drift, the system applies a three-stage output repair pipeline: (1)~strict JSON parsing for well-formed outputs, (2)~a bracket-balancing character-stack algorithm that appends missing closing delimiters to recover truncated arrays or objects, and (3)~pattern-based field extraction that recovers individual threat records by matching known schema field sequences, enabling partial recovery even when the model terminates mid-record.

Each threat card is further annotated with its phase within the MCP-UPD parasitic tool chain model~\cite{zhao2025mindserver}, which decomposes multi-tool attacks into three phases: parasitic ingestion (adversary embeds malicious instructions in external data), privacy collection (the agent collects sensitive data via legitimate tools), and privacy disclosure (exfiltration of collected data). Following Zhao et al.~\cite{zhao2025mindserver}, we label inter-tool transitions as T2T (tool-to-tool) and the final exfiltration step as UPD (unintended privacy disclosure). This annotation enables the knowledge graph to represent multi-hop attack paths as directed edges between tool nodes.

\subsubsection{Composite Risk Scoring}
\label{sec:arch-scoring}

To move beyond binary threat classification and support quantitative prioritization, MCPThreatHive includes a configurable composite risk scoring model. The platform ships with a default scoring formula adapted from the DREAD methodology~\cite{shostack2014threat}, reinterpreted for LLM-agentic contexts. Organizations can adjust the weights, factors, or replace the formula entirely to match their own risk appetite.

To align with established cybersecurity conventions, we employ a CVSS-style scale of 0 to 10 for risk scoring. The formula below computes a base risk score $R \in [0, 1]$ by combining four factors:
\begin{equation}
\label{eq:risk-score}
R = w_L \cdot L_{\text{norm}} + w_S \cdot S + w_I \cdot I + w_D \cdot D
\end{equation}
where the factors and default weights are listed in Table~\ref{tab:risk-factors}.

\begin{table}[tbp]
\caption{Example composite risk scoring factors and weights.}
\label{tab:risk-factors}
\centering
\small
\begin{tabular}{L{3.5cm} c c L{5.5cm}}
\toprule
\textbf{Factor} & \textbf{Symbol} & \textbf{Weight} & \textbf{Definition} \\
\midrule
Security Impact Severity & $L_{\text{norm}}$ & 0.35 & Normalized 7-level scale: $L/7$ \\
Attack Success Rate      & $S$               & 0.30 & Empirical exploitation probability $\in [0,1]$ \\
Persistence / Scope      & $I$               & 0.20 & Transient\,=\,0.5, Session\,=\,0.75, Long-term\,=\,1.0 \\
Exploitation Ease        & $D$               & 0.15 & Insider-only\,=\,0.33, Moderate\,=\,0.66, Trivial\,=\,1.0 \\
\bottomrule
\end{tabular}
\end{table}

A set of MCP-specific priority multipliers amplifies the score for threat classes that are unique to agentic systems: semantic or inference-time attacks ($\times 1.20$), parasitic chaining or multi-tool amplification ($\times 1.15$), and low runtime observability ($\times 1.10$). The final composite risk score is $R_{\text{final}} = \min(10.0,\; (R \times P) \times 10)$, classified as Critical ($\geq 9.0$), High ($7.0 \leq R_{\text{final}} < 9.0$), Medium ($4.0 \leq R_{\text{final}} < 7.0$), or Low ($R_{\text{final}} < 4.0$).

As an illustration, MCP-19 (Direct Prompt Injection) receives $L{=}6$ (RCE possible; $L_{\text{norm}} = 6/7$), $S{=}0.85$, $I{=}0.75$ (session-scoped), $D{=}1.0$ (trivially exploitable), yielding $R = 0.300 + 0.255 + 0.150 + 0.150 = 0.855$. As a semantic attack, the inference-time multiplier ($P = 1.20$) applies, giving $R_{\text{final}} = \min(10.0,\; (0.855 \times 1.20) \times 10) = 10.0$, classified as \emph{Critical}. This aligns with empirical evidence that direct prompt injection is the most validated and universally exploitable MCP threat class~\cite{yang2025mcpsecbench, zhang2025mcpsecuritybenchmsb}.

\subsubsection{Knowledge Graph Construction}
\label{sec:arch-kg}

Threat cards and their associated entities are persisted in both a relational database and a \emph{neuro-symbolic knowledge graph}. The graph combines the precision of rule-based extraction with the semantic generality of LLM inference. Deterministic pattern matching extracts high-confidence entities such as CVE identifiers, CWE numbers, and known attack technique keywords. Probabilistic LLM extraction covers semantic entities that resist enumeration: Threat, Mitigation, Component, Technique, Asset, and Vulnerability concepts expressed in free-form prose.

The graph stores six node types (Intelligence Item, Threat Entity, MCP Threat ID, CVE Identifier, Tool, Mitigation) and five edge types (\textsc{describes}, \textsc{instances\_of}, \textsc{exploits}, \textsc{chains\_into}, \textsc{mitigated\_by}). The \textsc{chains\_into} relationship captures MCP-UPD parasitic tool chain paths, supporting queries such as ``find all tools reachable from a given entry point via documented attack chains.''

To prevent graph fragmentation from duplicate entities, a \emph{three-tier entity resolution} pipeline canonicalises incoming entities:
\begin{enumerate}
  \item \emph{Exact match} (case-insensitive string equality, $O(1)$ hash lookup).
  \item \emph{Jaccard similarity} on 3-gram character shingles. Given entity strings $A$ and $B$ with shingle sets $S_A$ and $S_B$:
  \begin{equation}
  \label{eq:jaccard}
  J(A, B) = \frac{|S_A \cap S_B|}{|S_A \cup S_B|}
  \end{equation}
  Entities with $J \geq 0.75$ are automatically merged. This threshold handles pluralisation, minor typos, and hyphenation variants.
  \item \emph{LLM verification} for ambiguous cases ($0.50 \leq J < 0.75$). The LLM is prompted: ``Are `\{E\_new\}' and `\{E\_existing\}' the same security concept? YES or NO.'' This resolves semantically equivalent but lexically distinct entities (e.g., ``prompt injection'' vs.\ ``PI attack'').
\end{enumerate}

\subsubsection{Visualization and Risk Planning}
\label{sec:arch-vis}

The platform exposes four interactive views. The \emph{Threat Matrix} projects the derived 38 MCP threat patterns onto the standardized 4$\times$17 MCPSecBench attack surface grid~\cite{yang2025mcpsecbench}, rendering a unified heatmap where cell intensity encodes the sum of risk scores across all threats mapped to that cell. This grid highlights clustered risk areas across the four core MCP workflow phases (Task Planning, Tool Invocation, Response Handling, and Cross-Phase), offering administrators an intuitive overview of security postures. The \emph{3D Threat Landscape} maps the four MCPSecBench attack surfaces as a city-skyline metaphor: building height encodes the maximum risk score and building color encodes one of four attack surface categories derived from the MCP-38 taxonomy: Server APIs (blue --- authentication and RCE threats), Tool Metadata (green --- poisoning and supply-chain threats), Runtime Flow (red --- prompt injection and privilege escalation threats), and Transport (amber --- MitM, rebinding, and transport-layer threats). The \emph{Knowledge Graph Viewer} provides an interactive exploration of entity relationships and attack chains.

The \emph{AI Risk Planner} generates structured, per-threat risk plans using a Batch-Aggregate-Refine (BAR) strategy. Threats are processed in small batches; per-batch plans are generated independently, then aggregated. Duplicate mitigations are merged using Jaccard similarity. A final refinement pass produces the output document, which includes detection method recommendations, mitigation strategies with implementation priority, cross-framework references, and implementation effort estimates. Decomposing the task into batch, aggregate, and refine stages prevents the model from producing repetitive output across a large threat set.

\begin{figure*}[tbp]
\centering
\begin{subfigure}[b]{0.32\textwidth}
    \centering
    \includegraphics[width=\textwidth]{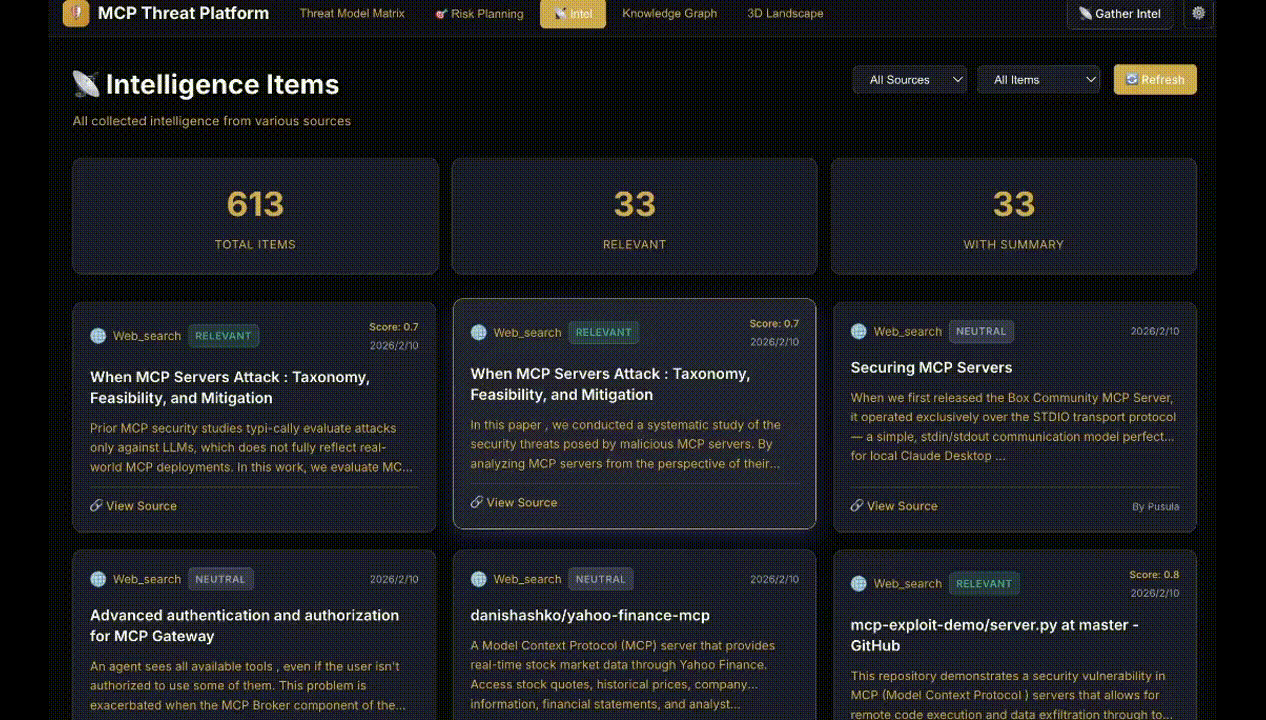}
    \caption{Intelligence Items}
    \label{fig:intel_items}
\end{subfigure}
\hfill
\begin{subfigure}[b]{0.32\textwidth}
    \centering
    \includegraphics[width=\textwidth]{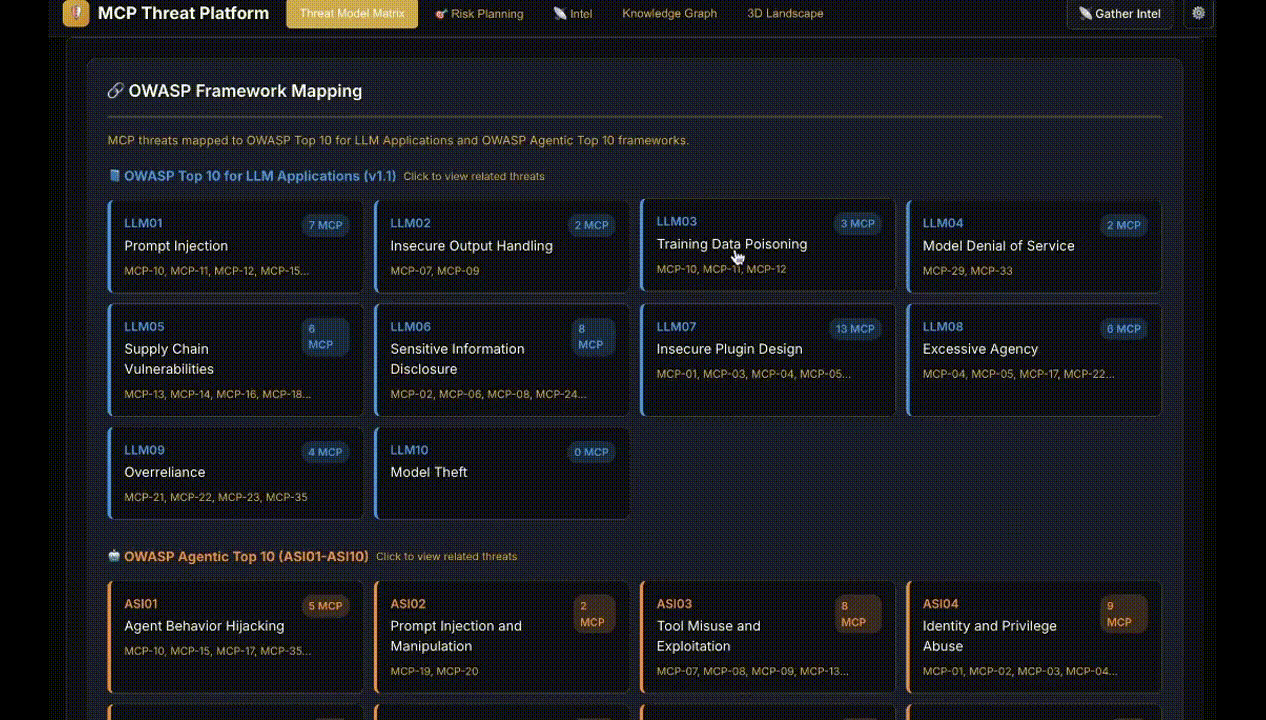}
    \caption{Threat Mapping}
    \label{fig:mapping1}
\end{subfigure}
\hfill
\begin{subfigure}[b]{0.32\textwidth}
    \centering
    \includegraphics[width=\textwidth]{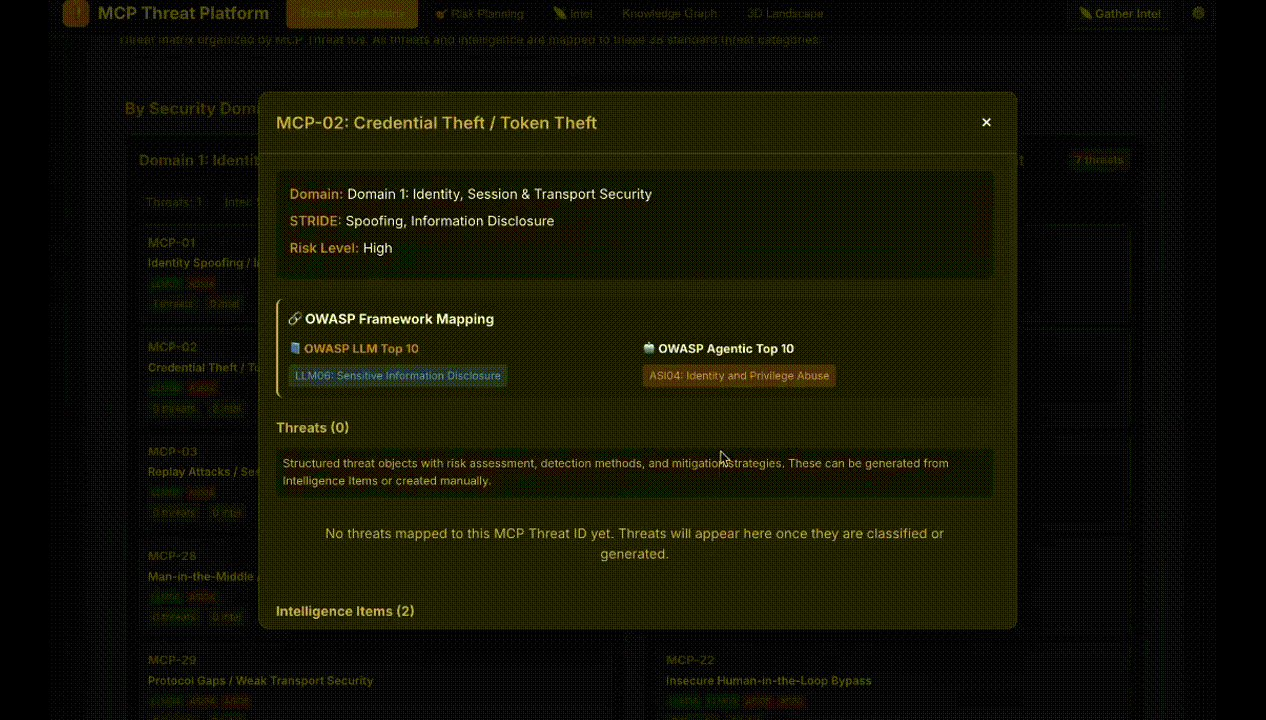}
    \caption{MCP-38 Description}
    \label{fig:mapping2}
\end{subfigure}

\vspace{0.5em}

\begin{subfigure}[b]{0.32\textwidth}
    \centering
    \includegraphics[width=\textwidth]{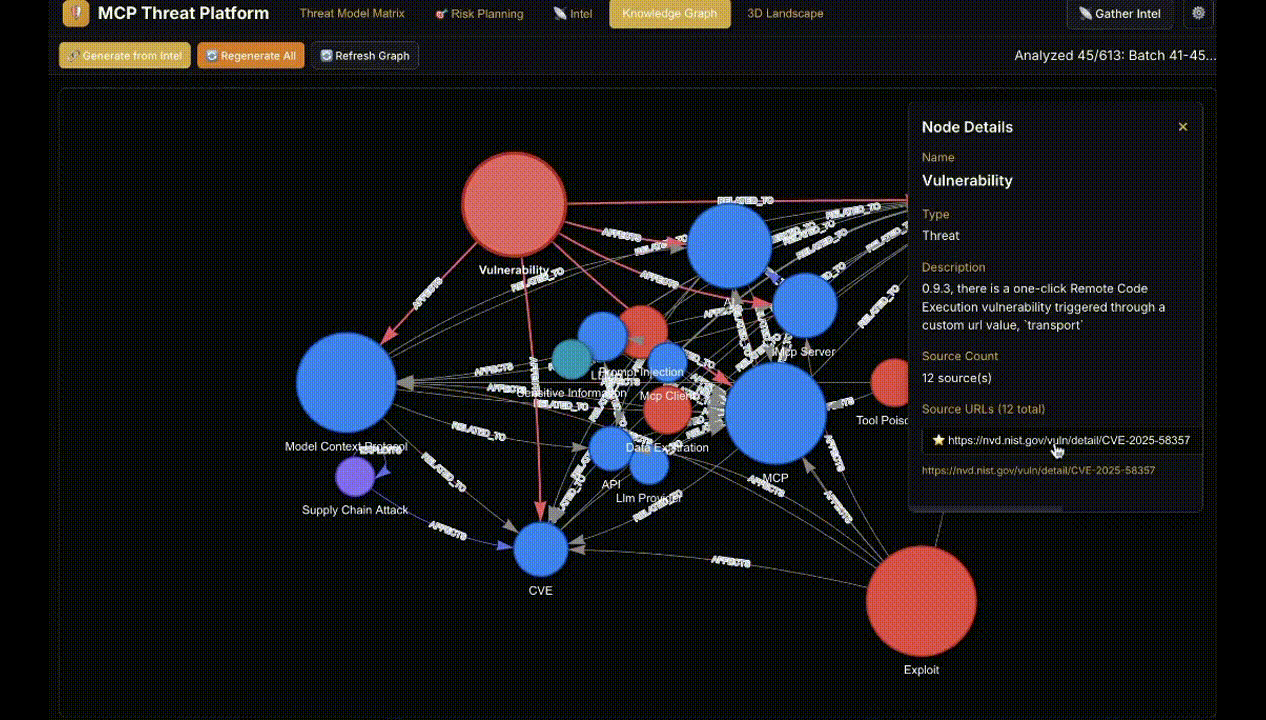}
    \caption{Threat Relationship Graph}
    \label{fig:neo4j}
\end{subfigure}
\hfill
\begin{subfigure}[b]{0.32\textwidth}
    \centering
    \includegraphics[width=\textwidth]{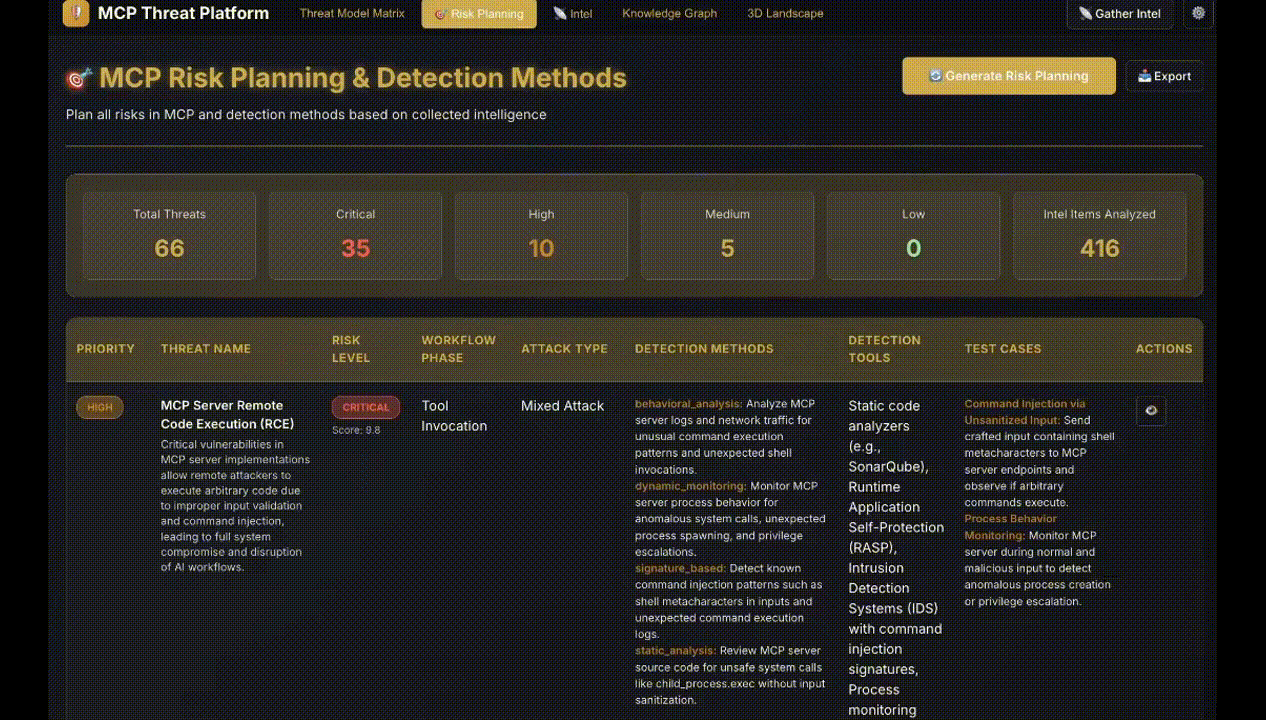}
    \caption{Risk Planning Dashboard}
    \label{fig:risk_planning}
\end{subfigure}
\hfill
\begin{subfigure}[b]{0.32\textwidth}
    \centering
    \includegraphics[width=\textwidth]{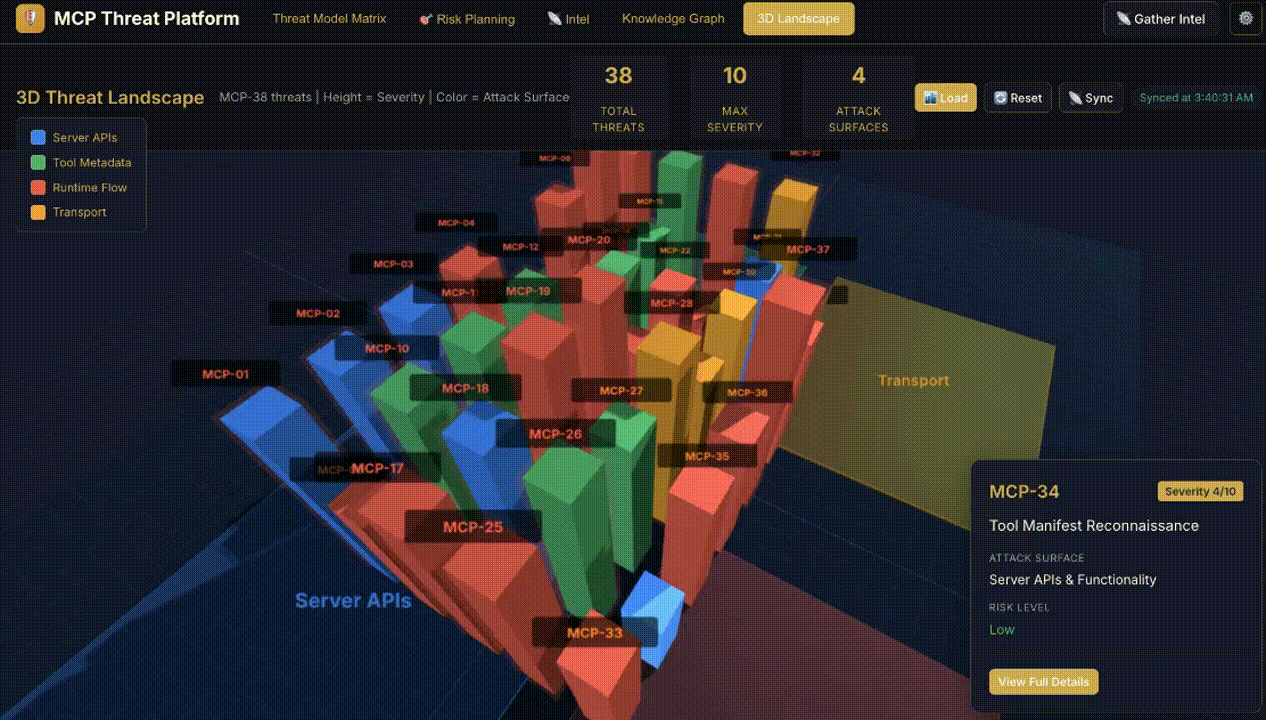}
    \caption{3D Threat Visualization}
    \label{fig:3d}
\end{subfigure}
\caption{MCPThreatHive system interfaces: (a)~intelligence gathering with relevance scoring, (b)~cross-framework threat mapping, (c)~per-threat detail card, (d)~knowledge graph visualization, (e)~risk planning dashboard, and (f)~3D threat landscape.}
\label{fig:system_interfaces}
\end{figure*}

\subsection{Implementation}
\label{sec:implementation}

MCPThreatHive is implemented in Python and released as open-source software.\footnote{\url{https://github.com/VulcanLab/MCPThreatHive}} The backend exposes a REST API via Flask with CORS support. Threat data is persisted in SQLite or PostgreSQL through SQLAlchemy ORM, while the knowledge graph is stored in Neo4j. LLM inference is provider-agnostic via LiteLLM~\cite{litellm2023}, supporting OpenAI, Anthropic, and local endpoints. The frontend uses JavaScript with Three.js for the 3D threat landscape visualization. The platform can be deployed via Docker Compose or run manually through an interactive setup wizard that configures LLM provider credentials and model selection. While the architecture is designed to support continuous threat intelligence monitoring, practical deployments initialise with a cold start phase: administrators manually trigger the intelligence gathering interface to ingest the baseline threat corpus and instantiate the foundational knowledge graph, after which subsequent updates can be processed incrementally.

Figure~\ref{fig:system_interfaces} presents the platform's key interfaces. The intelligence gathering view~(\ref{fig:intel_items}) displays collected items with relevance scores and source metadata. The threat mapping view~(\ref{fig:mapping1}) shows cross-framework mappings to OWASP LLM Top~10 and OWASP Agentic Top~10, while individual threat cards~(\ref{fig:mapping2}) expose per-threat detail including domain, STRIDE classification, and linked intelligence items. The knowledge graph~(\ref{fig:neo4j}) visualizes threat relationships in Neo4j, enabling analysts to trace multi-hop attack paths. The risk planning dashboard~(\ref{fig:risk_planning}) presents prioritized threats with detection methods, tooling recommendations, and test cases. Finally, the 3D threat landscape~(\ref{fig:3d}) renders all 38~MCP threats as a navigable cityscape where bar height encodes severity and color encodes attack surface. The dashboard additionally features a dedicated STRIDE Analysis view, rendering a distribution chart that segments classified threats across the six STRIDE categories (Spoofing, Tampering, Repudiation, Information Disclosure, Denial of Service, Elevation of Privilege), enabling security teams to rapidly identify which attack paradigm dominates the current intelligence corpus.


\section{Evaluation}
\label{sec:evaluation}

\subsection{Comparison with Existing Tools}
\label{sec:eval-comparison}

Existing MCP security tools fall into three categories: manifest scanners (e.g., Ramparts, Agentic Radar), runtime proxies (e.g., MCP-Guardian, Pangea MCP Proxy), and attack simulation platforms (e.g., MCPSecBench~\cite{yang2025mcpsecbench}, MCPSafetyScanner). Some tools span categories; MCP-Scan~\cite{mcp-scan-2025} combines static manifest analysis with runtime proxy interception. Table~\ref{tab:tool-comparison} compares MCPThreatHive with representative tools across key capability dimensions.

\begin{table}[tbp]
\centering
\caption{Capability comparison of MCP security tools. \checkmark = supported, ($\sim$) = partial, -- = not supported.}
\label{tab:tool-comparison}
\small
\begin{adjustbox}{max width=\linewidth}
\begin{tabular}{L{3.8cm} C{1.1cm} C{1.1cm} C{1.1cm} C{1.1cm} C{1.3cm} C{1.5cm}}
\toprule
\textbf{Capability} & \textbf{MCP-Scan} & \textbf{Ramparts} & \textbf{Agentic Radar} & \textbf{MCP-Guardian} & \textbf{MCPSec-Bench} & \textbf{MCPThreat-Hive} \\
\midrule
MCP-38 taxonomy coverage        & --          & --          & --          & --          & --          & \cellcolor{vulcanyellow!30}\checkmark \\
Continuous threat intelligence   & --          & --          & --          & --          & --          & \cellcolor{vulcanyellow!30}\checkmark \\
Knowledge graph construction     & --          & --          & --          & --          & --          & \cellcolor{vulcanyellow!30}\checkmark \\
AI-generated risk plans          & --          & --          & --          & --          & --          & \cellcolor{vulcanyellow!30}\checkmark \\
Multi-framework mapping          & --          & --          & ($\sim$)    & --          & --          & \cellcolor{vulcanyellow!30}\checkmark \\
MCP-UPD chain analysis           & --          & --          & --          & --          & --          & \cellcolor{vulcanyellow!30}\checkmark \\
Static manifest analysis         & \checkmark  & \checkmark  & \checkmark  & --          & --          & \cellcolor{vulcanyellow!30}($\sim$)   \\
Runtime proxy interception       & \checkmark  & --          & --          & \checkmark  & --          & \cellcolor{vulcanyellow!30}--          \\
Human-in-the-loop approval       & --          & --          & --          & \checkmark  & --          & \cellcolor{vulcanyellow!30}--          \\
Attack simulation                & --          & --          & ($\sim$)    & --          & \checkmark  & \cellcolor{vulcanyellow!30}--          \\
\bottomrule
\end{tabular}
\end{adjustbox}
\end{table}

The comparison reveals three gaps. First, existing tools model threats in isolation and cannot detect compositional attacks such as preference manipulation~\cite{wang2025mpma} and parasitic tool chains~\cite{zhao2025mindserver}, where the risk emerges from the interaction between individually benign tools. Second, all reviewed tools operate in manual or point-in-time mode rather than continuously monitoring intelligence sources. Third, per-tool scanners classify threats against a single framework (e.g., STRIDE or OWASP LLM) and do not provide unified multi-framework coverage spanning STRIDE, OWASP LLM, OWASP Agentic, and MCP-38. MCPThreatHive is the only tool that provides continuous threat intelligence, knowledge graph construction, AI-generated risk plans, and MCP-UPD chain analysis. Conversely, it does not perform runtime proxy interception or attack simulation, which are addressed by complementary tools such as MCP-Scan~\cite{mcp-scan-2025} and MCPSecBench~\cite{yang2025mcpsecbench}.

\subsection{Case Study: GitHub MCP Prompt Injection}
\label{sec:eval-case-study}

To validate that MCPThreatHive produces correct classifications, we test the pipeline against a publicly known, expert-labeled incident: the GitHub MCP prompt injection vulnerability disclosed in 2025~\cite{github-mcp-vuln-2025}. This is not a claim that the platform discovered the incident; rather, we use it as ground truth to verify that each pipeline stage produces the expected output when the incident is ingested as an intelligence item.

In the GitHub MCP incident, an attacker placed a crafted file in a public repository containing hidden prompt injection instructions. When a Claude agent using the GitHub MCP server read the file, the injected instructions redirected the agent to access tokens from a separate private repository and exfiltrate them through a subsequent tool call. Security researchers classified this as Indirect Prompt Injection combined with Data Exfiltration~\cite{github-mcp-vuln-2025}.

We fed the corresponding Cybernews article into the MCPThreatHive pipeline and recorded the output at each stage. Table~\ref{tab:pipeline-trace} summarizes the trace.

\begin{table}[tbp]
\caption{Pipeline trace for the GitHub MCP prompt injection incident. Each row shows a pipeline stage, its processing action, and the output produced.}
\label{tab:pipeline-trace}
\centering
\small
\begin{tabular}{L{2.5cm} L{5.0cm} L{5.0cm}}
\toprule
\textbf{Stage} & \textbf{Processing} & \textbf{Output} \\
\midrule
Intel Collector & Query: ``GitHub MCP prompt injection private repository'' fetches Cybernews article & \texttt{IntelItem} with relevance $= 0.94$ \\
\addlinespace
Relevance Filter & AI relevance score $0.94 > \theta = 0.70$ & Item passed to threat analysis \\
\addlinespace
Threat Analyzer & CoT: ``external file content acted as indirect injection vector; second tool call exfiltrated tokens'' & MCP-20 (Indirect Prompt Injection), MCP-24 (Data Exfiltration); STRIDE: Information Disclosure; OWASP LLM 2025: LLM01, LLM02; OWASP Agentic: ASI01, ASI02 \\
\addlinespace
MCP-UPD Analyzer & Chain identified: GitHub read tool (T2T) $\rightarrow$ token exfiltration tool (UPD) & Parasitic chain: T2T $\rightarrow$ UPD \\
\addlinespace
Knowledge Graph & New nodes: Threat $\xrightarrow{\text{INSTANCES\_OF}}$ MCP-20; Threat $\xrightarrow{\text{CHAINS\_INTO}}$ MCP-24 & 4 nodes, 3 edges added \\
\bottomrule
\end{tabular}
\end{table}

The AI-assigned primary classification MCP-20 (Indirect Prompt Injection) and secondary classification MCP-24 (Data Exfiltration via Tool Output) match the expert-assigned labels for this incident. The MCP-UPD chain annotation (T2T $\rightarrow$ UPD) correctly captures the two-phase attack structure: the agent first reads malicious content through a legitimate tool (parasitic ingestion), then exfiltrates data through a second tool call (privacy disclosure). This trace demonstrates that the pipeline's chain-of-thought prompting and schema-aware constraints produce classifications consistent with expert analysis on a known incident.


\section{Discussion}
\label{sec:discussion}

\subsection{Limitations}
\label{sec:limitations}

MCPThreatHive relies on LLM-based classification, which introduces limitations the platform mitigates but cannot eliminate. First, hallucination: the self-verification chain-of-thought step and schema-aware consistency constraints (Section~\ref{sec:arch-analysis}) reduce but do not prevent incorrect risk-level assignments. Second, token budget sensitivity: frontier reasoning models require a minimum output budget ($\geq$12,000 tokens) to avoid truncating their internal deliberation, a failure mode discovered during development and addressed by allocating a fixed output budget of 12{,}000 tokens (Section~\ref{sec:arch-analysis}). Third, domain shift: non-English threat reports or non-standard terminology may be misclassified; multilingual prompting is a planned mitigation. Fourth, false positives and misclassification: the classifier may flag legitimately described tools that use aggressive language (e.g., penetration-testing utilities) as malicious, mislabeling them as Sandbox Escape or Data Exfiltration threats. MCPThreatHive is intentionally designed as a high-confidence indicator system rather than an automated remediation engine; its outputs should be reviewed by a qualified analyst before any restrictive countermeasures are applied to deployed MCP server configurations.

The evaluation in Section~\ref{sec:evaluation} demonstrates classification fidelity on a single known incident. A formal empirical evaluation measuring precision and recall across a larger labeled incident corpus is needed to quantify classification accuracy at scale.

\subsection{Deployment Context}
\label{sec:deployment}

MCPThreatHive is designed for three primary audiences: (1)~security teams responsible for MCP-integrated AI products, using the threat matrix and risk planner to prioritize remediation; (2)~researchers studying MCP security, using the knowledge graph to identify threat chains and coverage gaps; and (3)~compliance teams using the cross-framework OWASP and STRIDE mappings to support security posture reporting.

The platform is a threat intelligence and modeling tool, not a runtime protection proxy or static manifest scanner. It complements runtime tools such as MCP-Scan~\cite{mcp-scan-2025} by providing the upstream intelligence layer: identifying and classifying threats that runtime scanners can then detect in deployed configurations.

\subsection{Ethical Considerations}
\label{sec:ethics}

The MCP-38 taxonomy and the threat cards generated by the platform describe attack techniques in sufficient detail to be useful for defenders. This dual-use risk is inherent to threat intelligence: the same information that helps a security team build detection rules could, in principle, inform an attacker. We mitigate this by focusing platform output on detection methods, mitigations, and risk prioritization rather than step-by-step exploitation procedures. The open-source release enables community review of the taxonomy and the AI classification logic, supporting transparency and reproducibility.


\section{Conclusion and Future Work}
\label{sec:conclusion}

We have presented MCPThreatHive, an open-source threat intelligence platform for Model Context Protocol agent ecosystems. The platform implements a four-stage pipeline (intelligence gathering, AI-powered threat analysis, knowledge graph construction, and interactive visualization) that continuously monitors multiple sources and automatically classifies emerging threats against the MCP-38 taxonomy~\cite{shen2026mcp38}, STRIDE, and the OWASP LLM and Agentic Top~10 frameworks. A composite risk scoring model adapted from DREAD provides quantitative prioritization, and a batch-aggregate-refine risk planner generates actionable mitigation strategies. A case study on the GitHub MCP prompt injection incident demonstrates that the pipeline produces classifications consistent with expert analysis.

Several directions remain for future work. First, a formal empirical evaluation of classification precision and recall on a larger labeled incident corpus would strengthen confidence in the pipeline's accuracy. Second, exporting platform intelligence as Semgrep or YARA rules would enable push-to-scanner threat feeds for runtime integration. Third, STIX/TAXII export would support cross-organisation threat intelligence sharing. Finally, continuous monitoring of MCP registries such as Smithery~\cite{smithery2025} and Glama~\cite{glama2025} would enable automated detection of new types of attacks in the wild.


\bibliographystyle{IEEEtran}
\bibliography{ref}

\end{document}